\documentclass[twoside,twocolumn,9pt]{article}
\usepackage{extsizes}
\usepackage[super,sort&compress,comma]{natbib} 
\usepackage[version=3]{mhchem}
\usepackage[left=1.5cm, right=1.5cm, top=1.785cm, bottom=2.0cm]{geometry}
\usepackage{balance}
\usepackage{times,mathptmx}
\usepackage{sectsty}
\usepackage{graphicx} 
\usepackage{lastpage}
\usepackage[format=plain,justification=justified,singlelinecheck=false,font={stretch=1.125,small,sf},labelfont=bf,labelsep=space]{caption}
\usepackage{float}
\usepackage{fancyhdr}
\usepackage{fnpos}
\usepackage[english]{babel}
\usepackage{array}
\usepackage{droidsans}
\usepackage{charter}
\usepackage[T1]{fontenc}
\usepackage[usenames,dvipsnames]{xcolor}
\usepackage{setspace}
\usepackage[compact]{titlesec}
\usepackage[utf8]{inputenc}

\usepackage{epstopdf}%This line makes .eps figures into .pdf - please comment out if not required.

\definecolor{cream}{RGB}{222,217,201}

\newcommand{\rvec}{ \mathbf{r} }
\newcommand{\Rvec}{ \mathbf{R} }
\newcommand{\fvec}{ \mathbf{f} }

\newcommand{\nvec}{ \mathbf{n} }
\newcommand{\vvec}{ \mathbf{v} }
\newcommand{\reff}{R^{\text{eff}}}

\begin{document}

\pagestyle{fancy}
\thispagestyle{plain}
\fancypagestyle{plain}{

%%%HEADER%%%
\fancyhead[C]{\includegraphics[width=18.5cm]{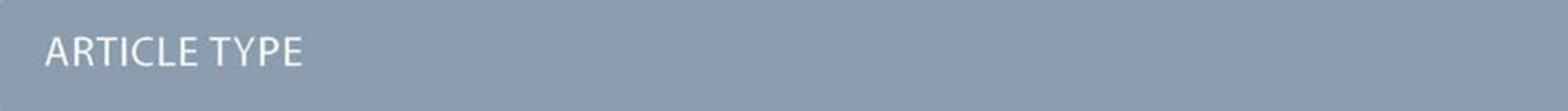}}
\fancyhead[L]{\hspace{0cm}\vspace{1.5cm}\includegraphics[height=30pt]{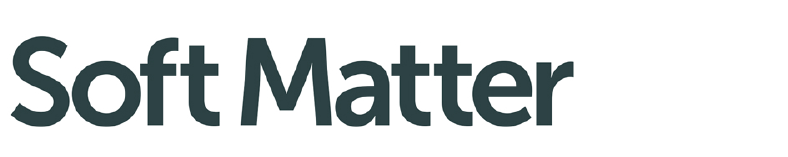}}
\fancyhead[R]{\hspace{0cm}\vspace{1.7cm}\includegraphics[height=55pt]{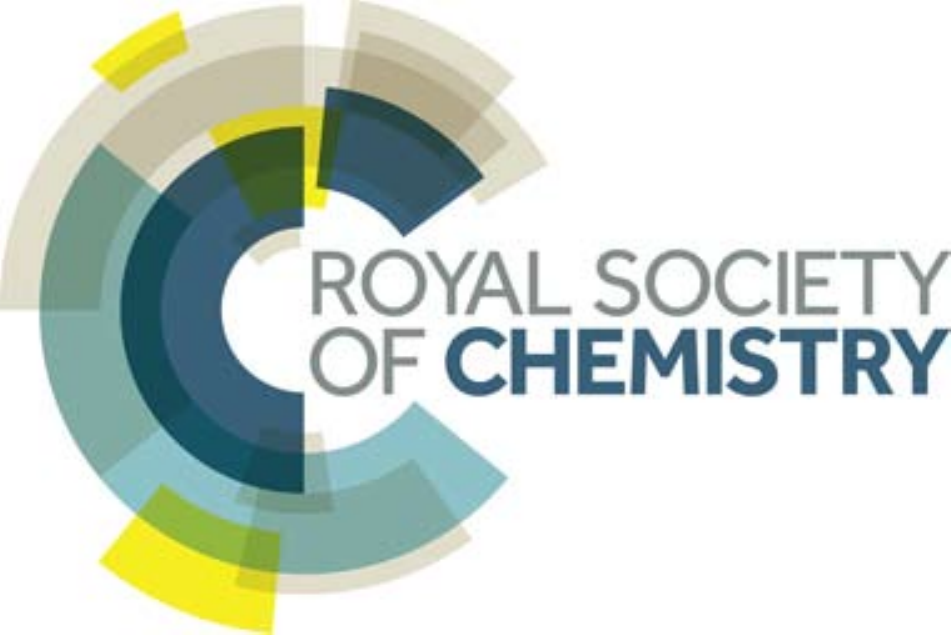}}
\renewcommand{\headrulewidth}{0pt}
}
%%%END OF HEADER%%%

%%%PAGE SETUP - Please do not change any commands within this section%%%
\makeFNbottom
\makeatletter
\renewcommand\LARGE{\@setfontsize\LARGE{15pt}{17}}
\renewcommand\Large{\@setfontsize\Large{12pt}{14}}
\renewcommand\large{\@setfontsize\large{10pt}{12}}
\renewcommand\footnotesize{\@setfontsize\footnotesize{7pt}{10}}
\makeatother

\renewcommand{\thefootnote}{\fnsymbol{footnote}}
\renewcommand\footnoterule{\vspace*{1pt}% 
\color{cream}\hrule width 3.5in height 0.4pt \color{black}\vspace*{5pt}} 
\setcounter{secnumdepth}{5}

\makeatletter 
\renewcommand\@biblabel[1]{#1}            
\renewcommand\@makefntext[1]% 
{\noindent\makebox[0pt][r]{\@thefnmark\,}#1}
\makeatother 
\renewcommand{\figurename}{\small{Fig.}~}
\sectionfont{\sffamily\Large}
\subsectionfont{\normalsize}
\subsubsectionfont{\bf}
\setstretch{1.125} %In particular, please do not alter this line.
\setlength{\skip\footins}{0.8cm}
\setlength{\footnotesep}{0.25cm}
\setlength{\jot}{10pt}
\titlespacing*{\section}{0pt}{4pt}{4pt}
\titlespacing*{\subsection}{0pt}{15pt}{1pt}
%%%END OF PAGE SETUP%%%

%%%FOOTER%%%
\fancyfoot{}
\fancyfoot[LO,RE]{\vspace{-7.1pt}\includegraphics[height=9pt]{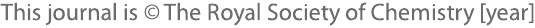}}
\fancyfoot[CO]{\vspace{-7.1pt}\hspace{13.2cm}\includegraphics{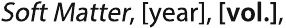}}
\fancyfoot[CE]{\vspace{-7.2pt}\hspace{-14.2cm}\includegraphics{RF}}
\fancyfoot[RO]{\footnotesize{\sffamily{1--\pageref{LastPage} ~\textbar  \hspace{2pt}\thepage}}}
\fancyfoot[LE]{\footnotesize{\sffamily{\thepage~\textbar\hspace{3.45cm} 1--\pageref{LastPage}}}}
\fancyhead{}
\renewcommand{\headrulewidth}{0pt} 
\renewcommand{\footrulewidth}{0pt}
\setlength{\arrayrulewidth}{1pt}
\setlength{\columnsep}{6.5mm}
\setlength\bibsep{1pt}
%%%END OF FOOTER%%%

%%%FIGURE SETUP - please do not change any commands within this section%%%
\makeatletter 
\newlength{\figrulesep} 
\setlength{\figrulesep}{0.5\textfloatsep} 

\newcommand{\topfigrule}{\vspace*{-1pt}% 
\noindent{\color{cream}\rule[-\figrulesep]{\columnwidth}{1.5pt}} }

\newcommand{\botfigrule}{\vspace*{-2pt}% 
\noindent{\color{cream}\rule[\figrulesep]{\columnwidth}{1.5pt}} }

\newcommand{\dblfigrule}{\vspace*{-1pt}% 
\noindent{\color{cream}\rule[-\figrulesep]{\textwidth}{1.5pt}} }

\makeatother
%%%END OF FIGURE SETUP%%%

%%%TITLE, AUTHORS AND ABSTRACT%%%
\twocolumn[
  \begin{@twocolumnfalse}
\vspace{3cm}
\sffamily
\begin{tabular}{m{4.5cm} p{13.5cm} }

\includegraphics{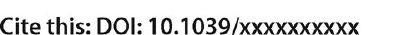} & \noindent\LARGE{\textbf{Co-assembly of Janus nanoparticles in block copolymer systems}} \\%Article title goes here instead of the text "This is the title"
\vspace{0.3cm} & \vspace{0.3cm} \\

 & \noindent\large{Javier Diaz\textit{$^{a}$}, Marco Pinna\textit{$^{a}$}, Andrei Zvelindovsky\textit{$^{a}$} and Ignacio Pagonabarraga \textit{$^{bc}$}} \\%Author names go here instead of "Full name", etc.

\includegraphics{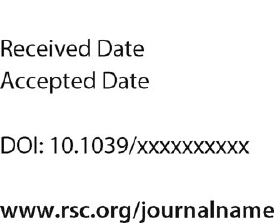} & \noindent\normalsize{Block copolymer are ideal matrices to control the localisation of colloids. Furthermore, anisotropic nanoparticles such as Janus nanoparticles possess an additional orientational degree of freedom that can play a crucial role in the formation of highly ordered materials made of block copolymers. This work presents a mesoscopic simulation method to assert the co-assembly of Janus nanoparticles in a block copolymer mixture, finding numerous instances of aggregation and formation of ordered configurations which can be related to dispersions of pure Janus colloids. Comparison with  chemically homogeneous neutral nanoparticles determines that Janus nanoparticles are less prone to induce bridging along lamellar domains, thus being a less destructive way to segregate nanoparticles at interfaces. The combination of asymmetric block copolymer and asymmetric Janus nanoparticles can result in assembly of colloids in an even number of layers within one of the block domains.} \\%The abstrast goes here instead of the text "The abstract should be..."

\end{tabular}

 \end{@twocolumnfalse} \vspace{0.6cm}

  ]
%%%END OF TITLE, AUTHORS AND ABSTRACT%%%

%%%FONT SETUP - please do not change any commands within this section
\renewcommand*\rmdefault{bch}\normalfont\upshape
\rmfamily
\section*{}
\vspace{-1cm}

%%%FOOTNOTES%%%

\footnotetext{\textit{$^{a}$ Centre for Computational Physics, University of Lincoln. Brayford Pool, Lincoln, LN6 7TS, UK}}
\footnotetext{\textit{$^{b}$~Departament de Fisica de la Materia Condensada, Universitat de Barcelona, Marti i Franques 1, 08028 Barcelona, Spain }}
\footnotetext{\textit{$^{c}$CECAM, Centre Europ\'een de Calcul Atomique et Mol\'eculaire, \'Ecole Polytechnique F\'ed\'erale de Lausanne,
Batochime - Avenue Forel 2, 1015 Lausanne, Switzerland } }
%Please use \dag to cite the ESI in the main text of the article.
%If you article does not have ESI please remove the the \dag symbol from the title and the footnotetext below.
\footnotetext{\dag~Electronic Supplementary Information (ESI) available: [details of any supplementary information available should be included here]. See DOI: 10.1039/cXsm00000x/}
%additional addresses can be cited as above using the lower-case letters, c, d, e... If all authors are from the same address, no letter is required

\footnotetext{\ddag~Additional footnotes to the title and authors can be included \textit{e.g.}\ `Present address:' or `These authors contributed equally to this work' as above using the symbols: \ddag, \textsection, and \P. Please place the appropriate symbol next to the author's name and include a \texttt{\textbackslash footnotetext} entry in the the correct place in the list.}

%%%END OF FOOTNOTES%%%

%%%MAIN TEXT%%%%

% =============================================================
% =============================================================
% =============================================================
% =============================================================
% =============================================================
% =============================================================

	%\doublespacing

\section{Introduction}

Block copolymers (BCP) are perfect candidates to template the position of colloidal nanoparticles (NP), due to their inherent periodicity and well-defined microstructure\cite{wagner_two-dimensionally_2017,bockstaller_block_2005},  which allows  nanoparticles to segregate to specific regions of the microphase-separated mixture. 
The enthalpic interaction between the colloidal surface and the surrounding polymer dictates which phase of the block copolymer is energetically more favourable for the particle, as a first approximation\cite{bockstaller_size-selective_2003}. Furthermore, the orientational alignment of shape-anisotropic nanoparticles within the block copolymer can be controlled by tuning the NP length scales with respect to the block copolymer domain spacing. 
For instance, Krook et al recently could control the orientation of nanoplates dispersed in a  lamellar-forming block copolymer\cite{krook_alignment_2018}. 
CdSe nanorods have been shown to undergo co-assembly with PS-b-PMMA to produce an ordered side-to-side configuration\cite{ploshnik_co-assembly_2010,ploshnik_hierarchical_2010}. 

Colloids with patchy surfaces have attracted considerable attention in the last decades\cite{walther_janus_2008,pawar_fabrication_2010,du_anisotropic_2011,walther_janus_2013} given their complex self-assembly processes. In particular, \textit{Janus} nanoparticles (JNPs) are colloids with two chemically distinct \textit{faces} or sides. 
The self-assembly of a system of JNPs is rich in morphologies\cite{ren_viscosity-dependent_2013}, for instance, Iwashita et al \cite{iwashita_density_2017} found different types of aggregation, along with alternating sheets of JNPs, which internally are organised in zigzag. Simulations have also explored the self-assembly of JNPs, finding several ordered phases such as lamellar-like using Brownian dynamics \cite{j.beltran-villegas_phase_2014,delacruz-araujo_shear-induced_2018}, 
Monte Carlo \cite{preisler_phase_2013} or Molecular dynamics \cite{miller_hierarchical_2009}. 

JNPs have been mixed with binary mixtures such as homopolymer blends, to find that domain growth is slowed by the segregation of JNP to the interface between domains\cite{nie_enthalpy-enhanced_2018,nie_synthesis_2016,bryson_using_2015}. Similar results have been found using Dissipative Particle Dynamics (DPD)  \cite{huang_effect_2012,zhou_nanorods_2018} while the interfacial tension in the presence of Janus colloids was studied by C. Zhou et al \cite{zhou_dissipative_2016}.     

Experiments involving JNP and block copolymers are comparatively rare. 
Recently, Yang et al \cite{yang_design_2017} synthesized JNP such that each side has an affinity towards one of the blocks of a PS-b-P2VP block copolymer chain, in order to segregate them at  the interface between domains. 
Janus nanoparticles in block copolymers have been shown to posses a higher interfacial absorption energy compared to evenly coated nanoparticles\cite{yang_janus_2017}, which motivates its use to arrange nanoparticles at block copolymer interfaces.

Similarly to binary mixtures, JNP in block copolymer have been studied using DPD\cite{dong_chain-stiffness-induced_2015,chen_polymerization-induced_2017}. Furthermore, the combination of chemical-anisotropy and non-spherical shapes was studied by Yan et al \cite{yan_self-assembly_2010}, finding slowed-down timescale in the lamellar ordering. Using SCFT,  random copolymer  and mixed brushed coated colloids at block copolymer interfaces were compared, finding that nanoparticles coeated with a mixed brush were more effectively segregated at interfaces \cite{kim_positioning_2009}.  SCFT/DFT simulations were used to study JNP orientation at block copolymer interfaces. In that case, two spheres of different size and chemical affinity are considered\cite{wang_janus_2012}.  The ordering and positioning  of JNP in a lamellar-forming BCP have been investigated  \cite{osipov_orientational_2018} considering symmetric and asymmetric JNP made of two spheres. 

In this work we make use of a hybrid cell dynamic simulation/Brownian dynamics approach to study JNP dispersed in BCP mixtures. % creo que aqui falta mas (ver comentarios de ignacio )
This in-grid/off-grid method combines a continuous description of the block copolymer with a discrete description of  nanoparticles. 
This allows to simulate a considerably large number of BCP periods along with a large number of particles, in the order of hundreds. 
Contrary to previous works (namely DPD simulations), we do not create Janus nanoparticles as a cluster of isotropic nanoparticles, nor as a dumbbell-like combination of two spheres. 
On the contrary, each JNP is considered as an individual object with inhomogeneous chemical behaviour.  

The flexibility of the CDS/Brownian dynamics scheme allows us to study systems with considerable generality, for instance, without restrictions on the morphology of the block copolymer or the chemical composition of the JNP. For this reason, we will attempt to consider nanoparticles with very different surface chemistry, in contrast with previous works in which JNP were mainly compatible with either of the blocks of the BCP. We aim to provide a full description of the JNP co-assembly in diblock copolymer, making the connection with their chemically homogeneous counterparts. While the model is valid both in two (2D) and three dimensions (3D), we will restrict our simulations to 2D. 
This will allow to study considerably large systems over long time scales. 
While a degree of richness in the assembly of JNP aggregates is definitely lost in two dimensions, the basic elements of the JNP/BCP assembly are captured in 2D, namely, the combination of phase-separated domains with a given periodicity in the block copolymer, along with the orientational degree of freedom of each JNP.

\section{Model}

In this section, we introduce a hybrid cell dynamic simulation/ Brownian motion method in which the block copolymer is described by differences in concentration, while the nanoparticles are individually resolved. Therefore, this is a coarse-grained method that combines a continuous description of the block copolymer with a discrete consideration of each colloidal particle. The treatment of the BCP-colloidal coupling differs from previous works \cite{pinna_modeling_2011,diaz_cell_2017,diaz_phase_2018} while the two-face character of the JNP is introduced in a way that is different from previous approaches \cite{krekhov_periodic_2013,guo_phase_2017}, mainly because our coupling free energy will be shown to involve a volume (surface, in 2D) integral, as opposed to a surface ( 
line, in 2D) integral. 
% can we just remove all mentions to previous works altogether? they do not really add anything special.

The BCP is modelled by the order parameter $\psi(\rvec,t)$ which is related to the differences in the local monomer concentration $\phi_A(\rvec,t)$ and $\phi_B(\rvec,t)$ of block A and B, respectively, 
\begin{equation}
\psi(\rvec,t)=\phi_A(\rvec,t)-\phi_B(\rvec,t)+(1-2f_0)
\end{equation}
where the composition ratio $f_0=N_A/(N_A+N_B)$ quantifies the overall monomer fraction of monomers $N_A$ over the total amount of monomers in the system. $\psi(\rvec,t)$ is considered the local order parameter, which has a value $0$ for the disordered-or homogeneous- state and $|\psi|>0$ for microphase-separated regions. 

The time evolution of $\psi(\rvec,t)$ is dictated by the conservation of mass, resulting in the Cahn-Hilliard-Cook equation \cite{cahn_free_1959,cook_brownian_1970}
\begin{equation}
\frac{\partial\psi ( \rvec, t )}{\partial t}=
M \nabla^2 \left[
\frac{\delta F_{tot} [ \psi] }{ \delta \psi}
\right]+
\eta ( \rvec, t)
\label{eq:cahn}
\end{equation}
with $M$ being a mobility parameter and $\eta(\rvec,t)$ being a Gaussian  noise  that satisfies the Fluctuation-Dissipation theorem
\begin{equation}
\langle \eta(\rvec,t) \eta(\rvec',t')\rangle =
-k_B T M \nabla^2 \delta(\rvec-\rvec')
\delta(t-t')
\end{equation}
for which we have used the algorithm given by Ball\cite{ball_spinodal_1990}. $k_BT$ sets the thermal energy scale of the system. 

The total free energy functional, $F_{tot}$,  is decomposed into purely polymeric, coupling and intercolloidal free energy, respectively, 
\begin{equation}
F_{tot}=
F_{OK}+F_{cpl}+F_{cc}
\label{eq:Ftot}
\end{equation}
where the purely polymeric contribution $F_{OK}$ is the standard Ohta-Kawasaki free energy \cite{ohta_equilibrium_1986}. 
In fact, $F_{OK}=F_{sr}+F_{lr}$ can be decomposed in short ranged
\begin{equation}
\label{eq:Fshort}
F_{\text{sr}}[\psi]=\int d\rvec 
\left[ 
H(\psi)+\frac{1}{2} D |\nabla\psi|^2 
\right]
\end{equation}
and long-ranged contribution, 
\begin{equation}
\label{eq:Flong}
F_{lr}[\psi]=
\frac{1}{2} B\int d\rvec \int d\rvec'
G(\rvec,\rvec')\psi(\rvec)\psi(\rvec')
\end{equation}
with $G(\rvec,\rvec')$ satisfying $\nabla^2 G(\rvec,\rvec')=-\delta(\rvec-\rvec')$,i.e., the Green function for the Laplacian. 
This term accounts for the connectivity of the blocks which differentiates this model from binary mixtures. 

The local free energy can be written as  \cite{hamley_cell_2000}
\begin{equation}
H(\psi)=
\frac{1}{2}\tau'\psi^2 
+\frac{1}{3} v(1-2f_0)\psi^3 +\frac{1}{4} u \psi^4
\end{equation}
where $\tau'=-\tau+A(1-2f_0)^2$, $u$ and $v$ can be related to the molecular structure of the diblock copolymer chain \cite{ohta_equilibrium_1986}. 
The local free energy $H(\psi)$ possesses 2 minimum values, $\psi_{-}^{eq}$ and $\psi_{+}^{eq}$, which are the values that $\psi(\rvec,t) $ takes in the phase-separated domains and correspond to A-rich and B-rich domains.  
Parameter $D$ in Equation \ref{eq:Fshort} is related to the interface width, $\xi=\sqrt{D/\tau'}$, between domains and $B$ in Equation \ref{eq:Flong} to the periodicity of the system $H\propto 1/\sqrt{B}$.

Contrary to the block copolymer -which is described continuously- JNPs are individually resolved. 
A suspension of $N_p$ circular colloidal  nanoparticles at positions ${\Rvec_p}$ in the BCP is introduced by a coupling term in the free energy,  which takes a simple functional form
\begin{equation}
F_{cpl}[\psi,\{\Rvec_i,\phi_i\}]= 
\sum_{p=1}^{N_p}
\sigma\int d\rvec\ \psi_{c}\left(\rvec-\Rvec_p \right)
\left[\psi(\rvec,t)-\psi_0  (\phi_p)   \right]^2
\label{eq:fcpl}
\end{equation}
with $\sigma$ a parameter that controls the strength of the interaction.

$\psi_c(\rvec)$ is a function that is attached to each JNP accounting for the size and shape of the Janus nanoparticle.
We will use 
\begin{equation}
\psi_{c} (\rvec) = 
\exp\left[
1-\frac{1}{1-\left( \frac{| \rvec   |}{R_{eff}}  \right)^\alpha} 
\right]
\label{eq:psici}
\end{equation}
with $\psi_c(r>R_{eff})=0$ because it provides a monotonically decreasing function with a vanishing derivative at the cut-off $R_{eff}$. 
Furthermore, $R_{eff}$ serves as the soft-core radius of the particle, while a hard-core radius can be defined as the distance from the center at which $\psi_c(R_0)=1/2$, identifying the particle size, which results in $\reff = R_0  \left( 1+1/\ln 2    \right)^{1/\alpha} $. Therefore,  $\alpha$   controls the sharpness of the decay of $\psi_c(r)$.

\begin{figure}[h!]
\centering
\includegraphics[width=0.75\linewidth]{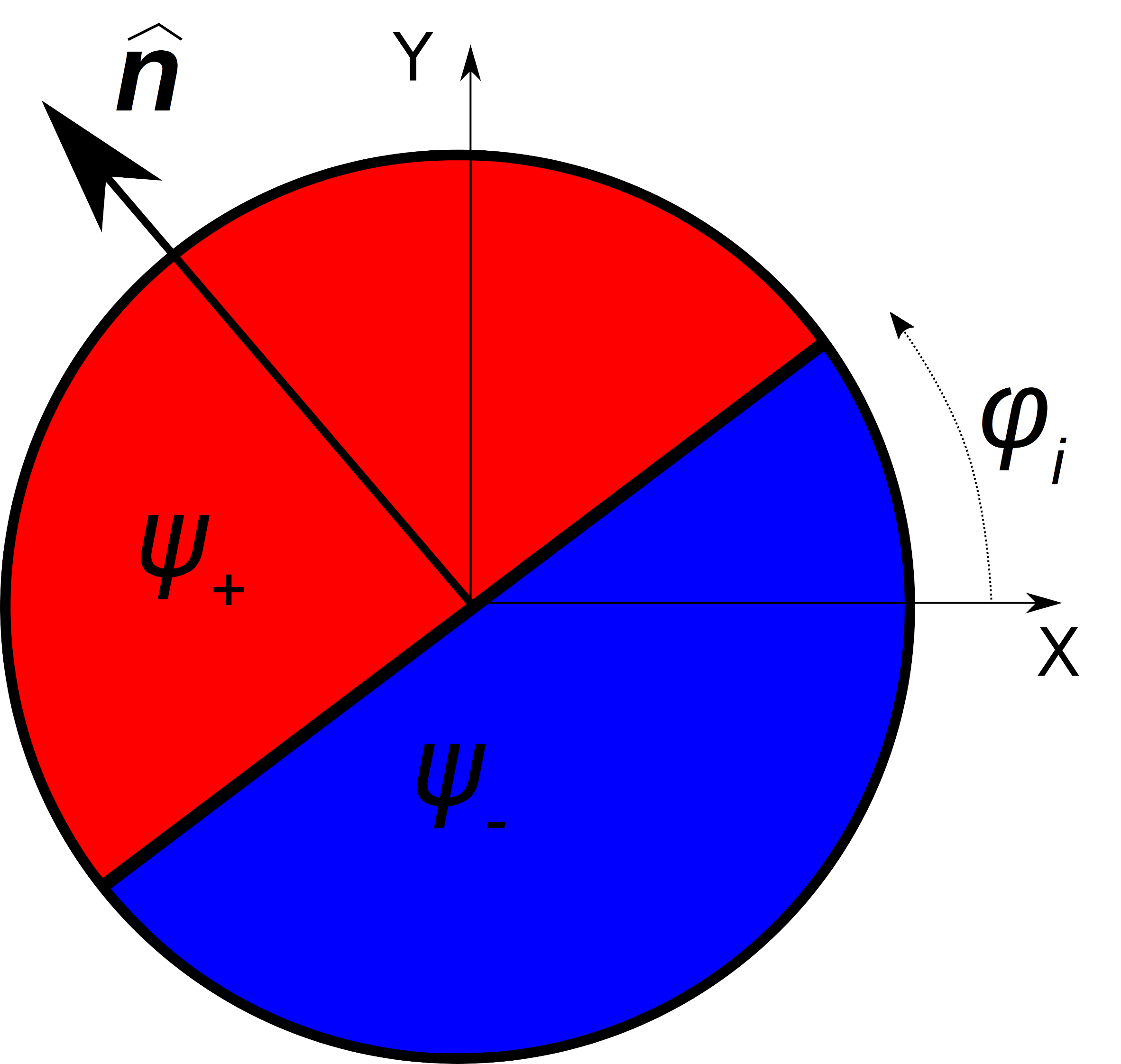}
\caption{Scheme of a modelled Janus nanoparticle. Two chemically distinct sides appear with regard to its chemical composition, with a unit vector $\nvec$ pointing into the positive half of the anisotropic nanoparticle.   }
\label{fig:scheme1}
\end{figure}

The affinity parameter $\psi_0(\phi_i)$ in equation \ref{eq:fcpl} represents the chemical properties of the NP. Contrary to chemically homogeneous nanoparticles, the coating of a JNP is anisotropic, taking two distinct values $\psi_0=\psi_+$ or $\psi_0=\psi_-$ for the negative and positive side of the JNP.
Each Janus nanoparticle has a unit vector $\nvec_i$ that controls the spatial distribution of chemical anisotropy, as in Fig.  \ref{fig:scheme1}, pointing into the positive side of the JNP. 
In order to characterize the JNP with generality it is useful to introduce two parameters
\begin{subequations}
\begin{equation}
\Delta\psi_0=\psi_+ - \psi_-
\label{eq:delta.def}
\end{equation}
\begin{equation}
\bar{\psi}_0=\frac{1}{2}\left(\psi_+ +\psi_- \right)  
\label{eq:mean.def}
\end{equation}
\label{eq:psi0.definitions}
\end{subequations}
where $\Delta\psi_0$ quantifies the chemical inhomogeneity of the particle, such that $\Delta\psi_0=0$ describes an homogeneous nanoparticle, while $\bar{\psi}_0$ describes the mean of the affinity of each side of the JNP, which is exactly equal to the affinity of a homogeneous colloid. These two parameters will help to characterize the JNP, while at the same time are useful to draw comparisons with homogeneous counterparts to a JNP. 

% ========================
% Np-NP interaction 
% ========================
The colloid-colloid interaction is introduced as a contribution to the total free energy $F_{tot}$ in equation \ref{eq:Ftot},
\begin{equation}
    F_{cc}=
    \sum_{i,j} V(\rvec_i-\rvec_j)
\end{equation}
where $V(r)$ describes non-overlapping, purely repulsive colloids. The interaction between colloids is chosen to be independent of the orientations of a pair of particles $\nvec_i$, $\nvec_j$ as the main intention of this work is to study the interactions between particles that are mediated by the BCP.
A Yukawa-like potential is chosen to describe the colloidal interaction
\begin{equation}
V(r)=
U_0 
\left[\frac{ 
\exp\left(1-r/R_{12}   \right)}{r/R_{12}}-1
\right]
\end{equation}
with $R_{12}=2R_0$ and $r$ being the center-to-center distance. 

 Colloids undergo diffusive dynamics, described by the Langevin equation in the over-damped regime. The center of mass of each colloid $\Rvec_i$ is considered to follow Brownian dynamics, that is,
\begin{equation}
\label{eq:brownian}
\vvec_i=
\frac{1}{\gamma} \left(
\fvec^{cc}+\fvec^{cpl}+\sqrt{2k_BT\gamma}\xi
 \right)
\end{equation}
with $\gamma$  the friction coefficient, $k_BT$ is the JNP thermal energy and $\xi$ is a random Gaussian term satisfying Fluctuation-Dissipation theorem. The coupling force $\fvec_i^{cpl}=-\nabla_i F_{cpl}$ accounts for the interaction between the nanoparticle and the BCP medium, while the colloid-colloid force arises due to the $F_{cc}$ contribution as in $\fvec_i^{cc}=-\nabla_i F_{cc}$.

%*****************************************
%     NUMERICAL IMPLEMENTATION
%*****************************************
The order parameter time evolution presented in Equation \ref{eq:cahn} is numerically solved using a cell dynamic simulation scheme\cite{oono_study_1988,bahiana_cell_1990} on a lattice, for which the laplacian is approximated as $\frac{1}{a_0^2}  [ \langle\langle X \rangle\rangle -X  ] $ with 
\begin{equation}
\langle\langle \psi \rangle\rangle = \frac{1}{6}  \sum_{NN}  \psi   +\frac{1}{12} \sum _{NNN} \psi
\end{equation}
in two-dimensional systems. NN and NNN stand for nearest-neighbour and next-nearest-neighbour, respectively, that is, summation over lattice points around the lattice point $\psi_{ij}$. 
The lattice is characterized by its spacing $a_0$. 

Initially, $\psi(\rvec,t=0)$ is randomly distributed, corresponding to a disordered phase. In addition, the initial state for the colloidal center of masses is a non-overlaping random distribution. The system is then let to evolve following the dynamical Equations \ref{eq:cahn} and \ref{eq:brownian} until a stationary state is approximately reached. 
Although a true equilibrium profile can not be assured, the time evolution of the  microphase separation of the diblock copolymer can be tracked\cite{ren_cell_2001} through $\langle | \psi(\rvec,t) | \rangle$ .    
A typical simulation run of a system  sized  $256\times 256$ requires a few hours of serial computational time. 

In summary, we use a cell dynamic simulation scheme coupled with a Brownian description of the time evolution of the particles. Each JNP interacts with the surrounding block copolymer via a shape function and a inhomogeneous affinity, which is split into two parts along the diagonal of the JNP. Each side  posses a different affinity value $\psi_+$ and $\psi_-$ which we will explore in the following sections.

\subsection{Relevant order parameters}

In order to quantify the different morphologies JNP and BCP give rise to, we will use a number of order parameters 

\textbf{Nematic order parameter.}

In order to characterize the local orientation of each colloid with respect to the block copolymer domains we consider the scalar product between the JNP orientation vector and the vector that is locally pointing normal to the interface of the BCP. 
\begin{equation}
S=< 2 (\mathbf{P} \cdot \nvec)^2-1 >
\label{eq:nematicorderparameter}
\end{equation}
with $\mathbf{P}\propto \nabla \psi$, that is, a normal vector that is normal to the interface between BCP domains. $S=1$ corresponds to a JNP  oriented normal to the interface and $S=-1$ when the particle is oriented along it. In the absence of net orientation, $S=0$.

\textbf{Inter-particle nematic order parameter.}

Similarly to the orientation between the BCP and the JNP, it is useful to consider the relative orientation between JNP, for example, when they form lamella-like arrangements. To this end, we consider
\begin{equation}
S_{inter-col}=<S^i_{inter-col}>_{i=1...N_p}; S_i=\sum_{r_{ij}<R^*}\left[ 2\left( \textbf{n}_i\cdot \textbf{n}_j \right)^2
-1 \right]
\end{equation}
in which we simply draw the same order parameter as in equation \ref{eq:nematicorderparameter} into the orientation of the $i$th particle with its closest neighbors given by a cut-off distance that we can typically set as $R^*=2.5 R_0$.

\textbf{In-cluster particle-to-center orientational order.}

Clusters of aggregating JNPs can be identified as a set of particles connected with a distance $r<d^*$ with $d^*$ determined from the radial distribution function. 
For a given cluster formed by  several nanoparticles, we can calculate the orientation of each particle with the centre of the cluster. This is different from analyzing the interparticle orientation as here we are interested on the scalar product of the orientation of a particle with the particle-to-center unit vector. To this end, we introduce 
\begin{equation}
Y_{orient}=<\nvec_i \cdot (\rvec_i-\rvec_{centre})>
\label{eq:in-cluster.orderparam}
\end{equation} 
with $<*>$ meaning an average over all particles in a cluster, and then averaging over all clusters (excluding single-particle clusters). $\rvec_i$ represents the position of a particle in a given cluster while $\nvec_i$ stands for its orientation. $\rvec_{center}$ is the geometric center of the cluster (which is carefully calculated according to the periodic boundary conditions).

\section{Results}

We aim to study in detail the co-assembly of Janus nanoparticles in block copolymer mixtures. In the simplest case, we expect the particles to simply be segregated towards their preferred region of the microphase-separated block copolymer. Nonetheless, simulations\cite{huh_thermodynamic_2000} and experiments\cite{halevi_co-assembly_2014,lee_effect_2002} have demonstrated that, at high  concentrations, the presence of nanoparticles can induce transitions in the block copolymer morphology. In the rest of this work, unless otherwise stated, we will use the standard set of parameters for the BCP 
$ \tau_0=0.35, u=0.5. v=1.5, A=1.5, D=1.0, B=0.002
$
while a radius $R_0=2.0$ will be used for the NP, in grid points. The lattice spacing $a_0$ is chosen to be unit. Furthermore, the temperature and the friction constant is generally fixed to be $T=1.0$ and $\eta_0=0.1$. 

\begin{figure}[h!]
\centering
\includegraphics[width=1.0\linewidth]{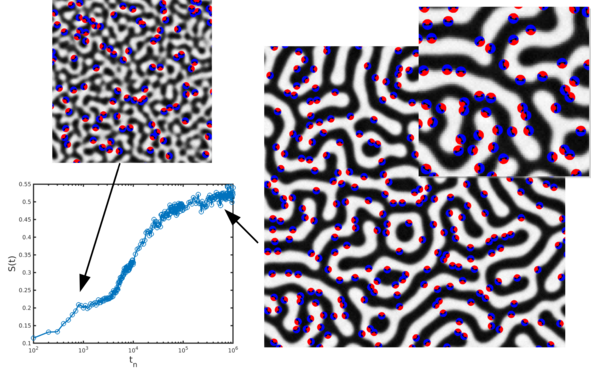}
\caption{Time evolution of a   JNP suspension in a symmetric diblok copolymer mixture. The orientation of the Janus particles, $S(t)$, is plotted against time, with two snapshots of early and late stages of the system.   }
\label{fig:example1}
\end{figure}

Firstly, we can assert the segregation and ordering of JNP in a simple symmetric diblock copolymer mixture with $f_0=1/2$. 
In Fig.  \ref{fig:example1} we can see the time evolution of a system of $N_p=300$ JNP with two antisymmetric sides, $\Delta\psi_0=1$ and $\bar{\psi}_0=0$, where the positive side has an affinity towards the white phase, and the negative has an exact same affinity towards the black phase.   At  $t_n=10^3$ (top snapshot) an early stage of BCP phase separation occurs, especially near a nanoparticle, which triggers phase separation in its vicinity. At $t_n=10^6$ (right snapshot) the lamella structure is well formed, while the JNP are all anchored at the interface between white and black domains. The orientation is always normal to the interface, as expected, and can be tracked in time with the curve of $S(t)$, where we can appreciate the orientational order parameter $S>0$ in the late stages of the simulation. Visually, we can observe that the $\psi_+$ red side of the JNP is facing the white domains. 

\begin{figure}[h!]
\centering
\includegraphics[width=1.0\linewidth]{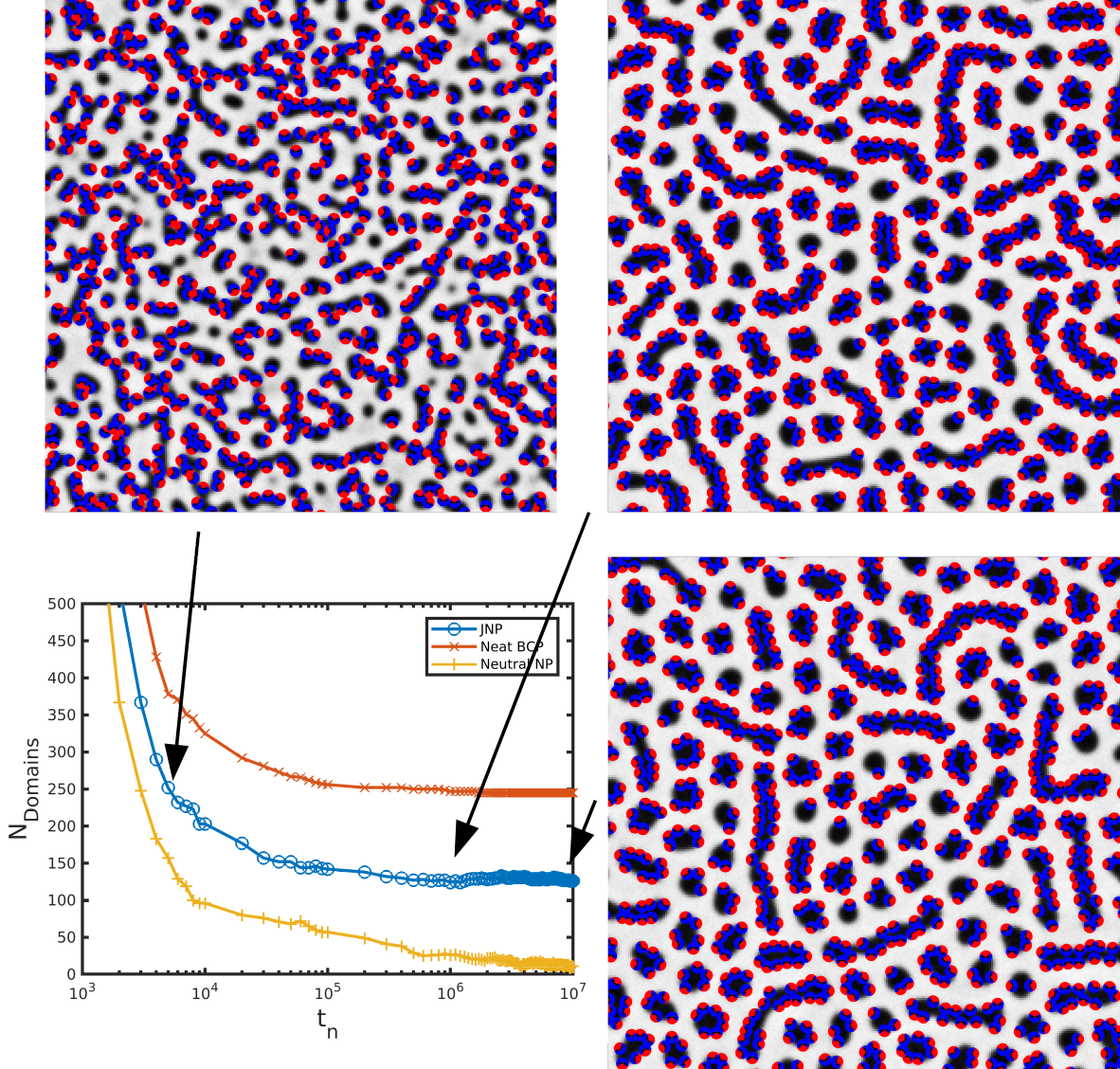}
\caption{Time evolution of the number of BCP domains, along with several images of the simulation results. The pure BCP is cylinder-forming ($f_0=0.35$) while the presence of the JNP favours domain merging.  }
\label{fig:example2}
\end{figure}

A cylinder-forming (circle-forming in 2D) asymmetric diblock copolymer mixture ($f_0=0.35$) can be mixed with $N_p=800$ Janus nanoparticles to assert the assembly of patchy particles at curved interfaces. 
The time evolution of an asymmetric BCP ($f_0=0.35$) /JNP mixture can be found in Fig.  \ref{fig:example2}. 
The decrease in the number of domains can be seen as time evolves, as expected for a cylinder-forming BCP. 
The JNPs are segregated to the interface and orient normal to it. 
We observe a coexistence of  circular-shaped BCP domains, with several black domains joined due to the presence of the JNP, resulting in elongated domains. 

Figure \ref{fig:example2} shows  the number of black domains as a functio of time, in the case of pure BCP ($N_p=0$) and homogeneous nanoparticles ($\Delta\psi_0=0$, $\bar{\psi}_0=0$). We can observe that the the number of black domains is reduced by the presence of JNP, but to a lesser extend than in the case of homogeneous  nanoparticles. In future sections we will study this behaviour in detail, but it serves as an introduction on the morphological changes induced by particles at interfaces, and the differences in the behaviour of Janus and chemically homogeneous nanoparticles.

\subsection{Phase diagram of BCP-JNP composites}

Generally speaking, a Janus nanoparticle does not interact equally with each block, for instance, it can be weakly attractive with one of the blocks while neutral to the other. 
This behaviour can be captured by the two affinities of the JNP, or, more suitably, by the differences and mean values of those affinities, $\Delta \psi_0$ and $\bar{\psi_0}$, respectively, as introduced in equation \ref{eq:psi0.definitions}. 
In this sense,  $\Delta\psi_0= 0$ describes a homogeneously coated nanoparticle, while  $\bar{\psi}_0$ describes the overall affinity of the colloid. 
We will analyze the parameter space in which the Janus properties are dominant.

%The Janus-like character of a particle is clearly given by $\Delta\psi_0$. Nonetheless, we can expect that small values in the anisotropy leads to a weakly anisotropic Janus particle that essentially behave like an isotropic particle, because the thermal motion dominates over the coupling interaction. 
%We can explore the parameter space given by $\Delta\psi_0 \text{vs} \bar{\psi}_0$ by simulating a moderate volume fraction of $\phi_p$ particles in a lamellar-forming diblock copolymer matrix ($f_0=0.5$). 

\begin{figure*}[h!]
\centering
\includegraphics[width=0.7\linewidth]{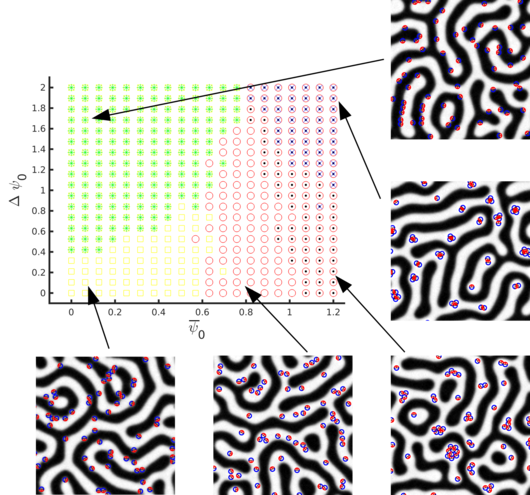}
\caption{Phase diagram of a small concentration of JNPs for different values of $\Delta\psi\ \text{vs}\ \bar{\psi}_0$, as defined in equations \ref{eq:delta.def} and \ref{eq:mean.def}. The Y axis determines the Janus-like property of the colloid (smaller values meaning weakly Janus) while the X axis determines the interface-compatible or A-compatible overall compatibility of the particle. 
 Yellow squares indicate nanoparticles which are anchored at the interface while red circles stand for NPs segregated within the domains. 
Yellow asterisks indicate orientational order of the colloids in the interface while black dots stand for nanoparticle aggregation. Finally, crosses indicate internal orientational order in the NP clusters. 
}
\label{fig:janus-phase1}
\end{figure*}

Figure \ref{fig:janus-phase1} shows the JNP assembled phases in a low colloidal concentration,$\phi_p=0.05$, such that the block copolymer assembly is weakly influenced by the presence of colloids.  
We vary the mean affinity $\bar{\psi}_0$ and the inhomogeneity parameter $\Delta \psi_0$. 

%In the $X$ axis the mean affinity $\bar{\psi}_0$ is explored 
%In the X axis the $\bar{\psi_0}$ parameter is explored. This means that the left-most part of the diagram is related to neutral-like particles while points on the right are expected to segregate to the center of domains. 
%The $Y$ axis describes the Janus-like characteristic of the colloids, such that in the $\Delta\psi_0 \sim 0$ region (bottom of the phase diagram), we observe the expected behaviour for an equivalent homogeneous colloid. 

At the bottom of the phase  diagram in Fig.  \ref{fig:janus-phase1},  we observe the expected behaviour, as values of the affinity which are close to zero lead to segregation of nanoparticles at the interface between domains (squares). 
As $\bar{\psi}_0>0$ increases, colloids are asymmetrically placed at the interface, up to a point when the colloids   detach from it and segregate towards the centre of the positive domains (circles).  
We calculate the distance $d$ from the JNP center of mass to the closest black domain to determine whether a particle is detached from the interface ($d>d_0 = 3.3$) or not. 
%The distance from the nanoparticle center of mass to a point $i,j$ in the grid such that $\psi<0.875$ indicates separation between a colloid and a black domain. We consider that a nanoparticle with a distance $d>3.3$ (in grid points) is detached from the interface. 
For values $\bar{\psi}_0>1$ the nanoparticles are not perfectly compatible with their hosting  domains and thus free energy minimization leads to aggregation of colloids (circles + dots). If the number of first neighbors is larger than $1$ we consider that a certain degree of aggregation is occurring.  
Negative values of $\bar{\psi}_0$ would show an equivalent behaviour in which colloids would segregate to the negative BCP domains. 

As we increase $\Delta\psi_0>0$ in Fig.  \ref{fig:janus-phase1}, the Janus-like nature of the particles becomes more dominant. When Janus particles have an antisymmetric positive/negative anisotropy (that is,when the two-fold affinity follows $\psi_{+}=-\psi_{-} \rightarrow \bar{\psi}_0= 0$ ) the JNPs segregate to the interface (left-most part of the phase diagram), and the torque orients them normal to the interface. 
This occurs as  Janus particles try to minimize the coupling free energy, which is satisfied when the positive side of the JNP is placed in the positive part of the phase-separated lamella. The combination of segregation to the interface and orientational order is marked in the phase diagram with an asterisk. Orientational order is characterised by the nematic-like order parameter, $S>0.5$, defined in equation \ref{eq:nematicorderparameter}. 

JNP ($\Delta\psi_0>>0$) with $\bar{\psi_0}>0$ are not symmetrically placed at the interface.  Instead,  we can expect that a small positive value in $\bar{\psi}_0$ results in a small displacement of the center of mass of the JNP into the A phase, therefore, an asymmetrical placement in the interface while maintaining an orientational order. This regime is denoted by asterisks in the left-most part of the phase diagram. 

As the displacement to the A domain becomes more prominent, the particle will  lose the orientational order (yellow squares) and eventually detach from the interface into the A-phase of the BCP. Nonetheless, the assembly of anisotropic nanoparticles in the A phase of the block copolymer is distinct from their homogeneous counterparts. Here, even if the overall coupling energy favours detachment from the interface, one of the sides of the Janus NP can be incompatible with the hosting BCP domain. One can consider, for example, a particle with a $\bar{\psi}_0=1$. For a chemically homogeneous particle,  both sides would have had the same affinity, $\psi_0=1$, and therefore it would be perfectly solvable into the A phase of the BCP. Instead, in a Janus NP both sides are introducing an energetic penalty that leads to distortions into the BCP profile. 
This distortion is minimized when particles form close-packed clusters. 
In turn, this leads to close-packed clusters of Janus NP even at relatively lower values of $\bar{\psi}_0$, which explains the shape of the dotted circle regime in the right part of the phase diagram.

In contrast to homogeneously coated nanoparticles, in Fig.  \ref{fig:janus-phase1} top-right snapshot, JNP form orientationally-ordered clusters. We can use the order parameter defined in Equation \ref{eq:in-cluster.orderparam} to quantify the in-cluster orientation. We consider that clusters with $Y_{\text{orient}}>0.7$ have an internal orientational order, as the normal vectors $\nvec_i$ point into the center of the cluster. 
These clusters acquire internal orientational order on top of a close-packed spatial organization. For instance, in Fig.  \ref{fig:janus-phase1} the positive (red) side  of the JNP is facing the interior of the cluster, which can be related to aggregation in self-assembly of patchy particles in solvents \cite{iwashita_density_2017,ren_viscosity-dependent_2013}

In summary, we have identified the regions of phase space in which the Janus nature of the particles are relevant but also the different and characteristic assemblies of Janus particles in block copolymer mixtures. This serves as an introduction to the rich phase assemble structures that can be formed in BCP-JNP composite system, in the limit of low concentration. 

\subsection{Comparison between Janus and chemically homogeneous (neutral) nanoparticles}

\begin{figure}[h!]
\centering
\includegraphics[width=1.0\linewidth]{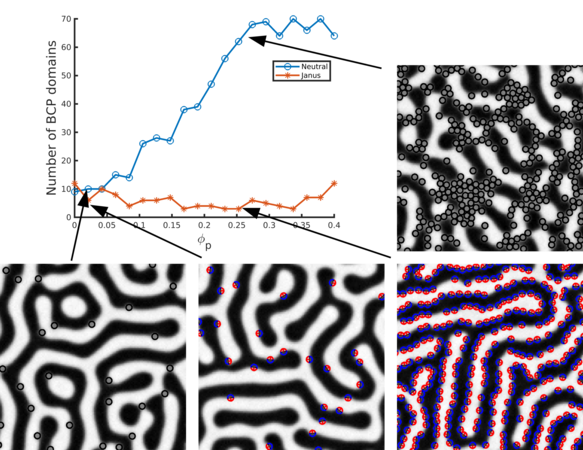}
\caption{Number of BCP domains as a function of the colloidal concentration.
The block copolymer is symmetric $f_0=1/2$. Homogeneous (neutral) and Janus nanoparticles are shown in different curves. }
\label{fig:janus-neutral}
\end{figure}

As we have seen in the previous section, Janus NPs have a strong tendency to segregate to the interface due to the inhomogeneous chemical coating on its surface, in which each side is compatible with one block. 
Chemically homogeneous($\Delta\psi_0=0$) neutral ($\bar{\psi}_0=0$) NPs also tend to segregate to the interface between domains as they are equally compatible with both blocks. 
Nonetheless, both types of assembly differ considerably, as JNP orient with $\mathbf{n}$ normal to the interface, while neutral nanoparticles are randomly oriented.
  We can compare neutral and Janus nanoparticles, as in Fig.  \ref{fig:janus-neutral}, where we simulate colloids in a symmetric diblock copolymer. 
Here, we find that at low concentration, both types of particles behave similarly, with neutral NPs segregating to the interface while only JNP also orient normal to the interface. The block copolymer is barely changed by the presence of a small fraction of particles, which can be tracked by the number of domains in the system. A low number of lamellar regimes is indicative of a well-ordered periodic structure. 
Nonetheless, we know from simulations and experiments that interface-segregated nanoparticles may form bridges across domains. 
This is a process of aggregation that may be undesirable if we want to have a well dispersed set of nanoparticles, to have an array of colloids, or even if destroying the lamella structure is inconvenient.

We can measure the degree of destruction of the lamella structure by calculating the number of different BCP domains present in the system. The lamella structure is characterized by having fewer domains than any other morphology. We can observe that neutral nanoparticles form clusters that lead to the formation of many small BCP domains. This does not occur with JNPs, which remain in the interface forming arrays. 

A similar analysis can be done regarding JNP and neutral NP in a cylinder-forming block copolymer. Fig.  1 in the Supplementary information shows that JNPs are less prone to induce unions of circular domains. In conclusion, JNPs are strongly anchored at BCP interfaces and also tend to protect the BCP morphology.  

\subsection{Orientational order vs thermal motion}

A JNP with two antisymmetric sides ($\bar{\psi_0}=0$) is expected to segregate to the interface and, if $\Delta \psi_0>0$, to orient with $\nvec$ pointing normal to the interface, into the positive region of the BCP, as was found in Fig.  \ref{fig:janus-phase1}. This is clearly not true for Homogeneous neutral particles, for which no orientational degree of freedom exists. We can also expect to find a regime of anisotropic particles in which the Janus-like nature is weak, therefore the Brownian noise is dominates. We have estimated this in the Supplementary Information (section 1), and obtained a parameter that controls the ratio between energetic coupling and thermal motion 
\begin{equation}
\chi= \frac{\Delta F_{cpl}}{k_BT} = 
\frac{16\ A_2}{\sqrt{2}}
\sigma R^2 \psi_{eq}^2
\frac{\psi_0}{\bar{\xi}\ k_BT}
\end{equation}  
where $\psi_0$ stands for the value of the affinity in the positive side of the JNP, given that the two-face JNP has $\psi_+=\psi_0$ and $\psi_-=-\psi_0$. $\psi_{eq}$ is a the equilibrium value of the BCP order parameter, that is, the (absolute) value of the BCP profile in the bulk. $A_2$ is a parameter related to the shape of the tagged function $\psi_c$ that is defined in the Supplementary Information Section 1. 

\begin{figure}[h!]
\centering
\includegraphics[width=1.0\linewidth]{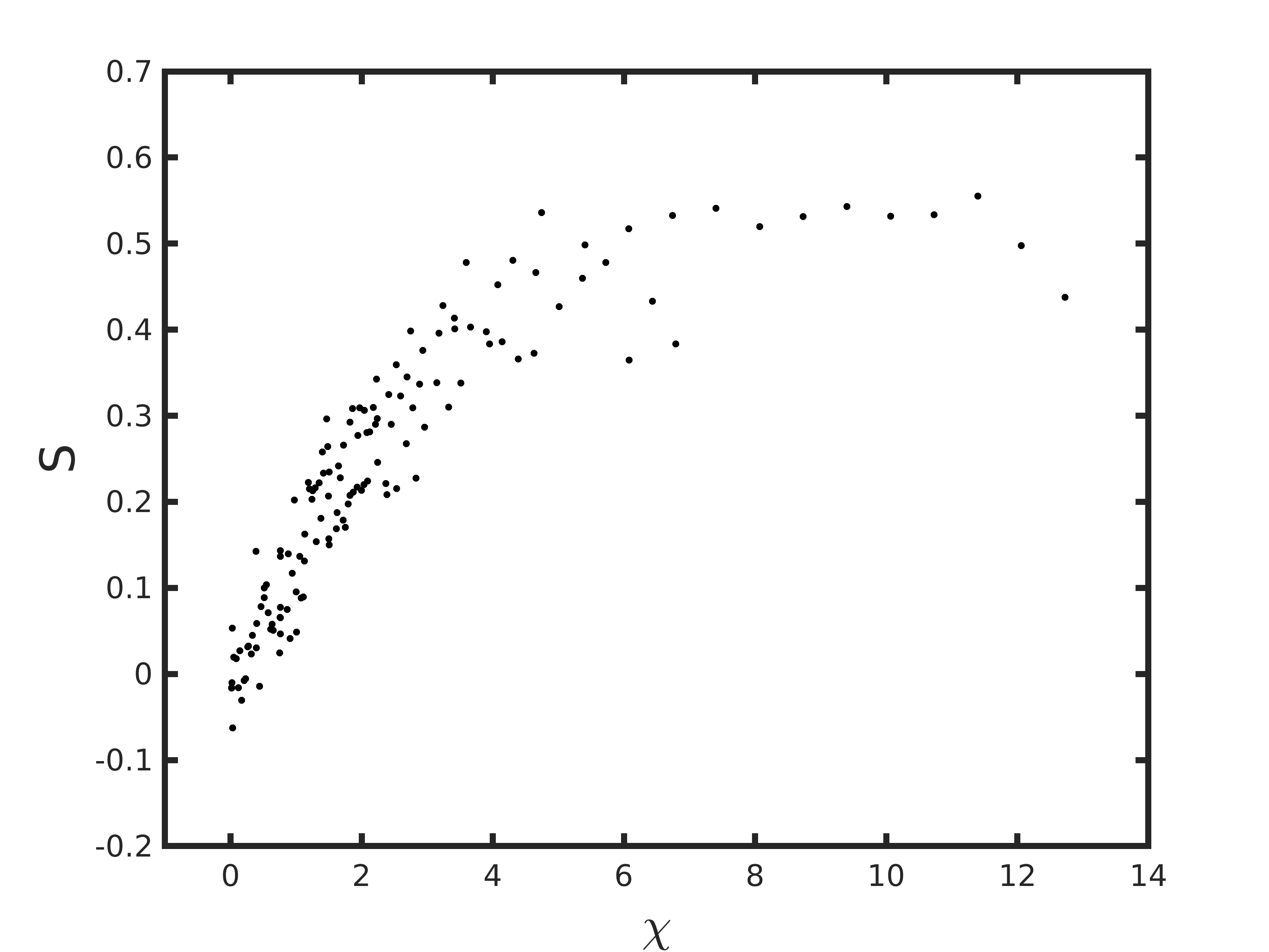}
\caption{Orientational order parameter $S$ of the JNP with respect to the BCP interface. Each point represents a different value of $\psi_0$ and $k_BT$ resulting in an ordered ($S>0$) or disordered ($S\sim 0$) configuration. }
\label{fig:S-chi}
\end{figure}

We can compare this with simulations, exploring different values of the temperature ($0.1<k_BT<5$) and different degrees of anisotropy ($0.01<\Delta \psi_0<5$). Fig.  \ref{fig:S-chi} shows the curve of the orientational order parameter $S$ with a single parameter $\chi$. We can see that the points approximately align  into a single curve. The lack of total order at $\chi>>1$ is explained as the assembly can lead to imperfections such as curved domains and defects in the BCP. 
The behavior clearly shows that for $\chi<1$ the thermal component of the Brownian motion is dominant, and so we find that disorder dominates $S\sim 0$. As opposed to the right-most region, in which the coupling is much stronger than the random fluctuation ($S>0$). These results can be related to SCFT/DFT simulations  by Wang et al \cite{wang_janus_2012}, where the order parameter $S$ grows monotonically with the size of the JNP. 

\subsection{Co-assembly of JNP in BCP}

In previous sections we have mostly considered a lamellar-forming symmetric block copolymer by fixing $f_0=1/2$. Nonetheless, we can consider the effect of Janus NP in a general case, exploring the composition parameter $f_0$ to assert the effect that a JNP with a fixed chemistry will affect the equilibrium profile of a BCP/JNP system. Thus, we are studying the complementary case to Fig.  \ref{fig:janus-phase1}, where for a fixed block copolymer we asserted the different JNP chemical surfaces.  

\begin{figure*}[h!]
\centering
\includegraphics[width=0.7\textwidth]{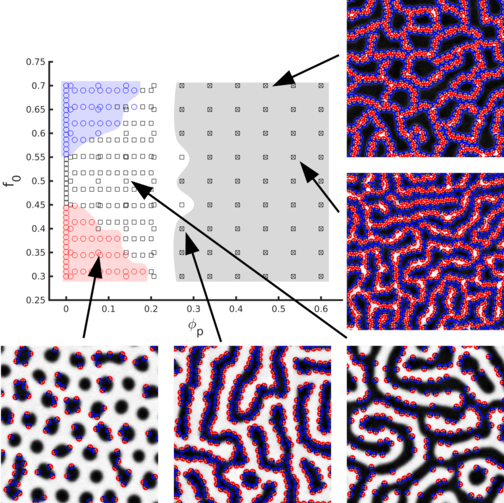}
\caption{
Phase behaviour of the BCP as a function of a JNP concentration $\phi_p$. The morphology of the  BCP is changed along the Y axis through the composition ratio of the chain $f_0$. The JNP here are symmetrical $\bar{\psi}_0=0$. Symbols describe the BCP morphology as:  blue circles $\circ$  for cylindrical phase with white block as minority phase;
  red circles $\circ$  for cylindrical phase with black block as minority phase;
   black squares for lamellar phase. 
  Furthermore, a black cross $\times$ represents points in which the interparticle orientational order parameter is larger than $0.1$.  
  }
\label{fig:phd.phip-f0.sym}
\end{figure*}

 The BCP morphology can be studied by  exploring the A to B monomer fraction $f_0$, while the importance of the JNP is characterized by the fraction of particles present in the system $\phi_p$, as shown in Fig.  \ref{fig:phd.phip-f0.sym}. The phase of the BCP system is characterised as circular domains (circle) or lamellar (squares). 
 We have performed  additional simulations without particles to establish the reference phase-behaviour of the block copolymer.
At low $\phi_p$ the colloids are simply segregated to the interface, as we have seen previously. The phase behaviour of the BCP is in this case dictated by the polymeric properties of the system. 

At high JNP concentrations, the presence of particles in  the interface leads to the union of different circular BCP domains if the BCP is indeed in the cylindrical phase ($|f_0-1/2|>0.05$). 
Due to this increase in the interface, a circular-to-lamellar phase is observed (blue circles to squares). 
At even higher surface fractions, the lamellar  domains are broken to accommodate a larger number of colloids into newly created interfaces. 
If the number of particles is increased, the BCP intrinsic order is destroyed, thus the system becomes totally occupied by a single percolating array of JNP.
Due to an effective interaction mediated by the surrounding block copolymer, the JNPs tend to orient side-to-side, with vectors $\mathbf{n}_i \cdot \mathbf{n}_j\approx 1$ for two neighboring particles.  

Figure \ref{fig:phd.phip-f0.sym} has  shown the effect that perfectly antisymmetric ($\bar{\psi}_0=0$) JNPs produce on an arbitrary morphology of BCP. 
Nonetheless, we can extend this study to off-center Janus colloids. 
As we have seen in Fig.  \ref{fig:janus-phase1}, such colloids can be detached from the interface or form orientationally-ordered clusters. 
We can consider a rather extreme case $\Delta \psi_0=1.0$ and $\bar{\psi}_0=1.0$. 
Each side of the JNP has an affinity $\psi_-=0.5$ and $\psi_+=1.5$. The negative side (in fact in this case it is positive) is weakly compatible with the interface, while the positive side is strongly incompatible with all phases, but still preferentially compatible with the positive phase (because $\psi_+>1$). Simulations\cite{diaz_cell_2017} have shown  that this incompatibility is resolved by clustering of nanoparticles, which reproduced experimental results \cite{ploshnik_hierarchical_2013}.

%As figure \ref{fig:janus-phase1} demonstrates, a JNP does not need to behave equally regarding each side's compatible block. That is, we can consider anisotropic particles in which $|\psi_+|\neq |\psi_-|$. We are therefore moving right in the phase diagram given by figure \ref{fig:janus-phase1}. 

\begin{figure*}[h!]
\centering
\includegraphics[width=0.7\textwidth]{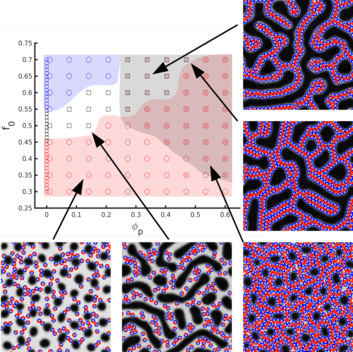}
\caption{Phase behaviour of the BCP in the presence of a concentration $\phi_p$ of Janus Nanoparticles with $\bar{\psi}_0=1.0$ and $\Delta \psi_0=1.0$. 
Symbols are as follows: Circles stand for cylindrical phase with color determining the majority (red circles indicate white monomer as the majority and blue circles indicate the opposite); lamellar phase is denoted by black squares; A further cross indicates a high degree of particle-to-particle orientational order.  
}
\label{fig:phd.phip-f0.asym}
\end{figure*}

In Fig.  \ref{fig:phd.phip-f0.asym} we study a range of $f_0$ values of the polymeric chain ratio and the fraction of particles in the system. Initially, at low particles concentration,  JNPs are simply segregated within the BCP without inducing any transformation in their hosting domains. In the bottom-left snapshot we can observe that JNP are weakly attached to the interface, with most particles dispersed in the positive BCP phase. Similarly, the 'positive' side of the JNP minimises the coupling free energy by creating clusters, but the $\psi_+$ value is not high enough to create more than a few weakly linked coupled particles. In both cases the thermal motion dominates.

In the region $f_0<0.45$, at high concentrations the nanoparticles are always occupying the majority phase. For this reason no phase transition is observed and the morphology of the BCP is always cylindrical ( red circles). Nonetheless, JNPs do undergo a disorder-to-order transition as the concentration is higher. A lamellar-like arrangement of  JNPs occurs, which is explained as  JNPs minimize the free energy by self-organizing as in Fig.  \ref{fig:phd.phip-f0.asym} bottom-right snapshot. Again, the thermal motion is also strong enough to avoid the formation of any long-range order of JNPs. 

Contrary to that, in the top-half of the phase diagram  ($f_0>0.5$) JNPs are compatible with the minority phase. Therefore, an additional effect of constrained has to be taken into consideration. Furthermore, JNPs are segregated into the positive BCP domains, therefore inducing a CYL $\rightarrow$ LAM phase transition (see the circle to square transition and the enhanced square region in the phase diagram). Because of this constrain,  JNPs are highly organized forming sheets or bilayer (snapshot top-left) or 4-layers (snapshot top-right). In both cases defects are strongly correlated with BCP defects.

\begin{figure}[h!]
\centering
\includegraphics[width=1.0\linewidth]{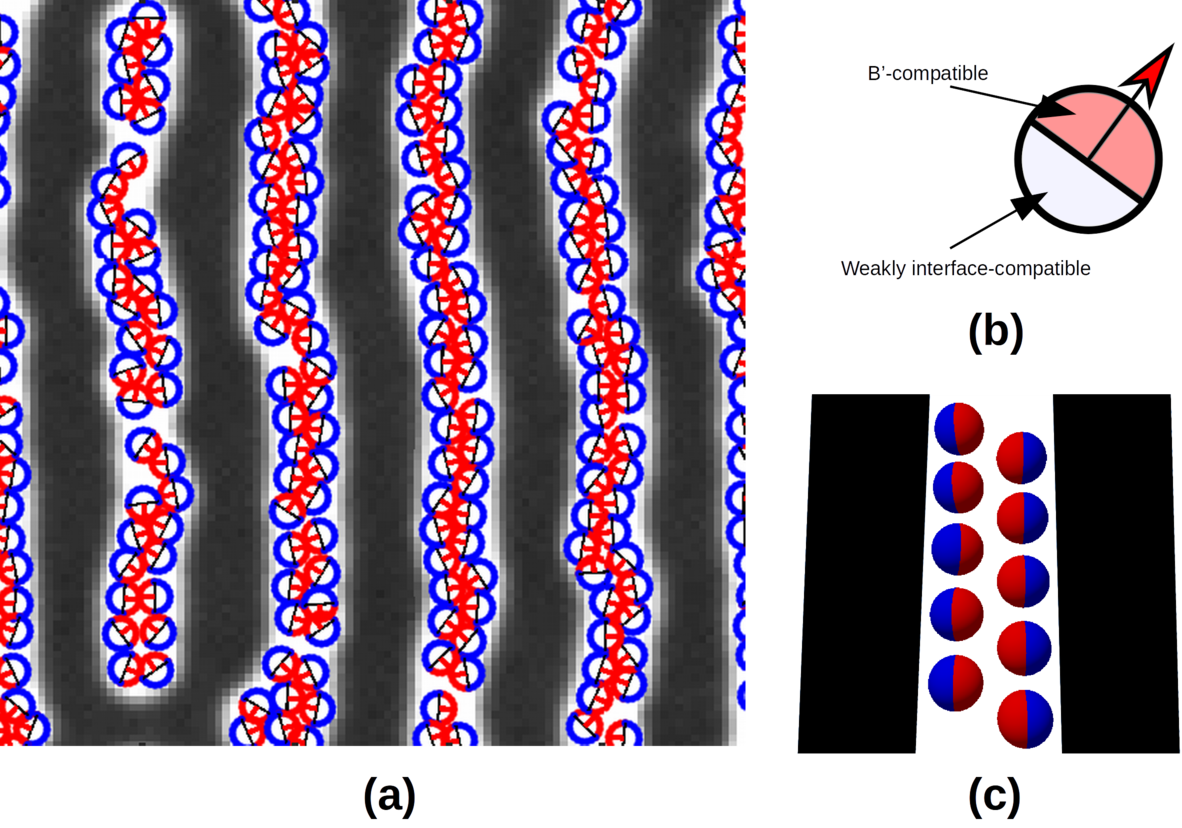}
\caption{ Co-assembly of an asymmetric diblock copolymer / asymmetric JNP mixture. The BCP undergoes a JNP-driven phase transition towards lamella due to the presence of the JNP in the interior of the white domains. The JNP form a bilayer due to their directional aggregation. Highly ordered simulation (a); scheme of the JNP (b) ; and scheme of the configuration (c).    }
\label{fig:newphase.scheme}
\end{figure}

In Fig.  \ref{fig:newphase.scheme} (a) we have simulated an initially-ordered BCP.
By doing so we can examine the long-range JNP order in the absence of curved interfaces.
JNPs are placed in the nodes of a triangular lattice, which suggests a close-packed arrangement of colloids within the white phase. 
In 3D, we would expect to observe a bilayer of JNPs in a 2D sheet with a thickness roughly corresponding to $4R_0$.  
In Fig.  \ref{fig:newphase.scheme}  (b) a simple scheme of the used JNP is shown while in Fig. \ref{fig:newphase.scheme} (c) the configuration is shown. 
This colloidal configuration is mediated by the block copolymer coupling that leads to highly ordered colloidal assembly. 
In summary, we have found a translational and orientationally ordered JNP configuration which is a consequence of BCP asymmetry ($f_0\neq 1/2$) and a precise choice in the JNP inhomogeneity. 
Indeed, the BCP undergoes a phase transition towards   a lamellar morphology, while the JNPs self-organize into lamellar-like arrangement.

\section{Conclusions}

We have used a hybrid cell dynamic simulation/Brownian dynamics method to study the dynamics and equilibrium properties of a system of Janus Nanoparticles in a diblock copolymer mixture. A combination of segregation, orientation and aggregation of particles within the mixture leads to a rich variety of assemblies, ranging from the expected interface-compatible antisymmetrical JNPs, to aggregation of colloids within one phase, in close resemblance with JNP suspended in a single solvent. In our study the Janus-like character of the particles has been explored ranging from completely Homogeneous particles, to particles with a clear two-face behaviour. 

JNPs have been shown to anchor at interfaces for a wider range of parameters than their homogeneous counterparts,  in accordance to previous results\cite{kim_positioning_2009}. The particles are found to orient normal to the interface between domains, with the A-compatible side pointing into the A-rich domain. Furthermore, JNPs are less prone to form bridges along BCP domains. These two properties are related, as JNPs are less likely to escape the interface and connect with other JNPs to create a broad, NP-rich area. This also means that the BCP structure is less affected by the presence of JNPs than with neutral nanoparticles. Therefore, using JNPs can be an effective way to segregate colloids at block copolymer interfaces, without inducing a break-up of lamellar-domains. 

For interface-compatible JNPs, the degree of orientation with respect to the block copolymer interface can be tracked and compared with the thermal fluctuations present in the Brownian motion. We have characterised the role of the temperature in the colloidal orientational order, to establish under which conditions the JNP orientation is dominated by its thermal random motion. This opens a way to control the properties of the JNP by tuning its thermal motion. As has been shown, the BCP lamellar texture is also strongly dependent on the anisotropy of the particle. 

Away from the interface, JNP are found to easily aggregate into clusters, as compared to homogeneously-coated colloids. This is due to one of the sides of JNP always being incompatible with its hosting phase. The BCP-induced aggregation can be related to instances of clusterisation of homogeneous nanoparticles within a incompatible environment \cite{ploshnik_hierarchical_2013}, while the organization is in close resemblance with several experiments and simulations of JNP in suspensions.  It is worth noing that we have not included any explicit  attractive interaction between colloids, nor does the interparticle potential include any angular dependence. Thus, this orientational aggregation is solely induced by the presence of the block copolymer.

Exploring the surface fraction of JNPs in BCP mixtures with arbitrary $f_0$ we have been able to determine different phase regions, as the presence of JNPs leads to transitions in the morphology of the BCP. 
Moreover, the morphology of the BCP also depends on the chemical properties of the JNP. 
While antisymmetric JNPs are segregated to the interface, a large amount of them can induce a transition in order to create a larger amount of interface in which they can anchor. 
Once this lamellar-to-cylindrical transition is completed, the JNPs dominate the morphology of the overall system, with JNP forming an almost totally connected network of colloids in an attempt to form maximize the interface. 

Asymmetric JNP  in BCP display a richer phase diagram. When one of the sides is compatible with the interface (neutral) and the other is strongly incompatible with one of the phases, the JNP have a weak tendency to form clusters within the less incompatible domain. When the fraction of particles is large the NPs  form a bilayer in order to minimize the BCP-JNP coupling. The parameter space of this type of assembly is given not only by the fraction of particles in the system, but also by the block copolymer morphology, as this assembly is driven by the level of constrain induced by the BCP. At even higher fraction of particles, the JNP are again dominating the co-assembly, and tend to form larger even number of layers, with $(1-3-5-..)$-layers being prohibited by the JNP two-face nature, along with the presence of the block copolymer. This lamellar-like organisation of JNP can be easily compared with sheet-forming assembly of Janus particles in solvents. At the same time, the organisation within the BCP (all nanoparticles are oriented normal to the interface) is similar to shape-anisotropic particles such as nanorods \cite{ploshnik_co-assembly_2010}. 

In summary, the co-assembly of JNP within block copolymer has been found to result in a large variety of highly-ordered configurations, both for the BCP and the colloids, in contrast to a mixture of chemically homogeneous colloids/BCP or a simple suspension of JNP in a fluid. The combination of JNP anisotropy and the inherent periodic morphologies of the diblock copolymer has proved to be essential to create new structures in which a precise control over the JNPs position and orientation can be achieved. 

\section{Conflicts of interest}
There are no conflicts to declare.

\section*{Acknowledgements}

J.D. would like to acknowledge the financial support of the BritishSpanish Society and Plastic Energy through the 2018 BSS-Plastic Energy Scholarship. J.D. would also like to acknowledge useful discussions with Dr Iwashita. 
IP acknowledges MINECO and DURSI for financial support
under projects FIS2015-67837-P and 2017SGR844,
respectively.

% =============================================================
% =============================================================
% =============================================================
% =============================================================
% =============================================================
% =============================================================
% =============================================================

%%%END OF MAIN TEXT%%%


\balance


%%%REFERENCES%%%
\bibliography{references} %You need to replace "rsc" on this line with the name of your .bib file
\bibliographystyle{rsc} %the RSC's .bst file

\end{document}